\documentclass[twocolumn,showpacs,preprintnumbers,amsmath,amssymb]{revtex4}

\usepackage{graphicx}
\usepackage{dcolumn}
\usepackage{bm}
\numberwithin{equation}{section}

\begin{document}
\author{Xiaohua Wu}\author{Yu Shaolan} \author{Zhou Tao}
\address{Department of Physics, Sichuan University, Chengdu 610064, China.}
\title{Designing  the unambiguous discriminator from the one-photon interferometer }

\begin{abstract}
 The quantum states filtering, whose general theorem
 was given by Bergou \emph{et al.} (Phys.Rev.A \textbf{71}, 042314(2005)),
  should find it's important
 applications in present scheme, where we  are trying to show that the
problem of quantum states unambiguous discrimination  may  be solved
by applying the argument of filtering. Let's use the quantum
filtering, as an example, to show the basic idea of present scheme.
Suppose there are N linearly independent states, if we are able to
find a (N+1)-dimensional
  unitary
 transformation, $R(\omega)$ ( with   $\omega$
 is an adjustable variable(s)), which will be performed on each
  $\vert\Psi>$ in the
way like: $R(\omega)\vert\Psi_1>=d_{11}(\omega)\vert D_1>+
f_{1}(\omega)\vert F>$,
 $R(\omega)\vert\Psi_k>=\sum_{l=2}^N d_{kl}(\omega)\vert D_l>+ f_{k}(\omega)\vert
 F>$,
then, according to the definition of the  operators for filtering,
there should be: $E_1(\omega)=R^{-1}(\omega)\vert D_1><D_1\vert R
(\omega)$, $E_2(\omega)=\sum_{k=2}^N R^{-1}(\omega)\vert
D_k><D_k\vert R (\omega)$ and $E_0(\omega)=R^{-1}(\omega)\vert
F><F \vert R (\omega)$. With this
 this $\{E_m(\omega)\}$ in hands, we could find the optimal
operators which lets  the function
$F(\omega)=\sum_{j=1}\eta_j<\Psi_j\vert E_0(\omega)\vert\Psi_j>$,
with $\eta_j$ to be  the ${\sl a~ priori}$ probability of
$\vert\Psi_j>$, have it's minimum value. For the system with N=3,
there are three types of operations: (a)
$E^a_1\vert\Psi_2>=E^a_1\vert\Psi_3>=0$, $ E^a_2\vert\Psi_1>=0$
and  $E^a_0$ corresponds to fail; (b) $E_i^b\vert\Psi_j>=0$ if
$i\neq j$ for i=1,2,3, and $E^b_0$ for failure; and (c)
$E^c_1\vert\Psi_2>=E^c_1\vert\Psi_3>=0,$
$E^c_2\vert\Psi_1>=E^c_1\vert\Psi_3>=0,$ while $E^c_0=\textbf{I }
-\sum_{k=1}^2E_k.$ We shall show that all these three types of
operators, which may be performed on a N=3 systems, can be get by
applying  argument of filtering: the case a is in fact the
filtering with N=3, case b can be viewed as successive filtering
and the case c can also be solved by an argument of filtering in
subspace. It can be shown that each  case, which belongs to the
above three, can be solved by reducing it to the problem of
filtering. An important case of N=4 system, has also been
discussed.
\end{abstract}


\maketitle

\section{introduction}
 As a very recent development, the possibility of unambiguous
discrimination between unknown quantum states can be potentially
useful for many applications in quantum computing and quantum
communications. The problem of unambiguous discriminating pure
states, which are successfully identified with nonunit probability
but witout error, was originally formulated and analyzed by
Ivanovic, Dicks and Peres [1-3] in 1987. Later, Jeager and Shimony
solved the question of unambiguous discrimination of two known pure
states with arbitrary ${ a~priori}$ probability. Shortly after this
result, Chefles proved that only linearly independent pure states
can be unambiguously discriminated [5]. The problem of
discrimination among three nonorthogonal states was first considered
by Peres and Terno [6], and the same question has also been
discussed by Duan and Guo[7] and Sun ${\sl et~ al.}$ [8]. Chefles
and Barnett also provided the optimal failure probability and it's
corresponding optimal measurement for a n symmetric states [9], and
an experimental set for discriminating four linearly independent
nonorthogonal symmetric states was given by Jim\'{e}nez ${\sl et~
al.}$ [10]. A new strategy for optimal unambiguous discrimination of
quantum states was also offered by Jafarizadeh ${\sl et~ al.}$ [11].

Unambiguous discrimination involving mixed state or a set of pure
states, became an object of research recently. Several necessary
and sufficient conditions for the optimum measurement have been
given by Zhang ${\sl et~ al.}$ [12]  and Eldar ${\sl et~ al.}$
[13]. Reduction theorems, which can simplify the discrimination
theorem, have been developed by Raynal ${\sl et~ al.}$ [14-15].
Low bounds for the failure and the conditions for saturating the
boumds, have also been studied [16-20]. There are only a few
special cases have analytical solution for the quantum
measurement, for examples, the quantum state filtering [21-23],
two mixtures with orthogonal or one-dimensional kernels [14-15],
two mixtures in The Jordan basis [24] and other cases [26-30).

In present work, we shall present a new scheme to solve the
problem of quantum  state  unambiguous discrimination. Let's use
the quantum state filtering originated from [21-23], as an
example, to show the basic idea of present scheme. Suppose there
are N linearly independent states, the task of the quantum state
filtering can be viewed as to find a set of operators $\{E_m\}$,
 whose elements are defined by : $E_2 \vert\Psi_1>=0$, $
 E_1\vert\Psi_k>=0$ with $2\le k \le N$, and $E_0$ corresponds to
 fail. If we are able to find a (N+1)-dimensional
  unitary
 transformation, $R(\omega)$ with $\omega$
 is an adjustable variable(s), which will be performed on each
  $\vert\Psi>$ in the
way like: $R(\omega)\vert\Psi_1>=d_{11}(\omega)\vert D_1>+
f_{1}(\omega)\vert F>$,
 $R(\omega)\vert\Psi_k>=\sum_{l=2}^N d_{kl}(\omega)\vert D_l>+ f_{k}(\omega)\vert
 F>$,
then, according to the definition of the  operators, there should
be: $E_1(\omega)=R^{-1}(\omega)\vert D_1><D_1\vert R (\omega)$,
$E_2(\omega)=\sum_{k=2}^N R^{-1}(\omega)\vert D_k><D_k\vert R
(\omega)$ and $E_0(\omega)=R^{-1}(\omega)\vert F><F \vert R
(\omega)$,
  With this $\{E_m(\omega)\}$ in hands, we could find the optimal
operators which lets  the function
$F(\omega)=\sum_{j=1}\eta_j<\Psi_j\vert E_0(\omega)\vert\Psi_j>$,
with $\eta_j$ to be  the ${\sl a~ priori}$ probability of
$\vert\Psi_j>$, have it's minimal value.

For the system with N=3, there are three types of operations: (a)
$E^a_1\vert\Psi_2>=E^a_1\vert\Psi_3>=0$, $ E^a_2\vert\Psi_1>=0$ and
$E^a_0$ corresponds to fail; (b) $E_i^b\vert\Psi_j>=0$ if $i\neq j$
for i=1,2,3, and $E^b_0$ for failure; and (c)
$E^c_1\vert\Psi_2>=E^c_1\vert\Psi_3>=0,$
$E^c_2\vert\Psi_1>=E^c_1\vert\Psi_3>=0,$ while $E^c_0=\textbf{I }
-\sum_{k=1}^2E_k.$ We shall show that all these three types of
operators, which may be performed on a N=3 systems, can be get by
applying  argument of filtering: the case a is in fact the filtering
with N=3, case b can be viewed as successive filtering and the case
c can also be solved by an argument of filtering in subspace. It
looks as if each  case, which belongs to the above three, can be
solved by reducing it to the problem of filtering. An important case
of N=4 system, has also been discussed.

Our present paper is organized as follows. Section II is a
preliminary section in which we introduce the so-called
double-triangle representation. In section III, we shall
 give a different way of solving the question of quantum states
 filtering. A concept of filtering in subspace will
 be introduced  in Sec.IV.   Two examples,
  discriminating three pure states and discriminating two mixtures
for  N=4, will be  discussed  in Section V and  VI, respectively. In
Sec.VII, we conclude the paper with a short summary.

\begin{figure}
\centering
\includegraphics{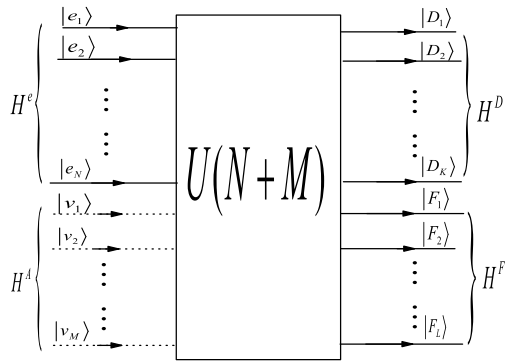}
\caption{\label{fig:epsart} An optical setting to realize the
POVMs in an enlarged Hilbert space}
\end{figure}
\section{double triangle representation}
\subsection{preliminary}
Considering a quantum system prepared in one of N pure quantum
states $\vert \Psi_j>$, where j=1, 2,..., N, if the states are
non-orthogonal, no quantum operations can deterministically
discriminate them. It is, however, possible to device a strategy
reveal the state with zero error probability under the condition
that these states are linearly independent [5]. Employing the Kraus
representation of quantum operations [31], each of the possible
distinguishable outcomes of an operation is associated with linear
transformation operators $\hat{A}_m$,
\begin{equation}
\sum_{m=o}^N\hat{A}^{\dagger}_m\hat{A}_m=\textbf{I},
\end{equation}
with $\hat{A}_0$ leads to failure while $\hat{A}_j$ corresponds to
the discrimination of $\vert\Psi_j>$. By introducing the ${\sl
reciprocal ~states}$ $\vert \Psi^{\bot}_j>$, which is defined as
that which lies in $ \mathcal{ H}$, the N-dimensional Hilbert
space for the N linearly independent states $\vert\Psi_j>$, and is
orthogonal to all $\Psi_{j'}$ for $j\neq j'$, Chefles found that
[5]
\begin{equation}
\hat{A}_j=\frac{p^{1/2}_j}{<\Psi^{\bot}_j\vert\Psi_j>}\vert
e_j><\Psi^{\bot}_j\vert,
\end{equation}
where $\vert e_j>$ form an orthonormal basis for $\mathcal{H }$
while $P_j$ is the conditional probability, given that the system
was prepared in the state $\vert\Psi_j>$, that this state will be
identified,
\begin{equation}
<\psi_j\vert\hat{A}^{\dagger}_{j'}\hat{A}_{j'}\vert\Psi_j>=P_j\delta_{jj'}.
\end{equation}
In the terms of positive operator valued measures (POVMs) [31],
the measurement
 can be expressed by defining the positive Hermitian operators
\begin{eqnarray}
E_j&=&\hat{A}^{\dagger}_{j}\hat{A}_{j}=\frac{p_j}{\vert<\psi^{\bot}_j\vert\psi_j>\vert^2}\vert\Psi_j^{\bot}>
<\Psi_j^{\bot}\vert,\nonumber\\
E_0&=&\hat{A}^+_{0}\hat{A}_{0},
\end{eqnarray}
with $E_0+\sum_{j=1}^N E_j=\textbf{I}$, and it has been shown that
the optimum measurement corresponds to the maximum eigenvalue of
value of $E_D=\sum_{j=1}^N E_j$ being equal to 1 [5].

Let $G\equiv\{\vert\Psi_j>\}$ for j=1, 2,..., N, the POVMs given
above can viewed as one type of operations  on G. There may be other
types of operations on the same G, for example, if there are two
known groups of states,
 $G_1\equiv\{\vert\Psi_k>\}$ for k=1, 2, ..., K, and
$G_2\equiv\{\vert\Psi_l>\}$ for l= L, L+1, ..., N, $G_1$ and $G_2$
may have common elements if $ L\le K$, we could also define a new
set of POVMs $\{E_i,E_0\}$ for i=1, 2,  $E_i$ can unambiguously tell
whether a state $\vert\Psi_?>\in G$ belongs to $G_i$ or not. Now,
one may ask the question: could  this $\{E_m\}$  also be expressed
in terms of $\vert\Psi^{\bot}_i><\Psi^{\bot}_j\vert$? We shall give
an answer to this question.  According to the Neumark's theorem
[32]: if each $E_m$ is an one dimensional positive operator, $\{E_m
\}$ can always be realized by extending the Hilbert space to a
larger space and performing orthogonal measurement in the larger
space, while, as we shall shown, we are able to realize the
discrimination of the quantum states just according to the
definition of the operators, this fact makes it possible to read
$\{E_m \}$ from their corresponding projective operators in the
enlarged space. We shall show how this basic idea works via the aid
of Fig.1: the total space is defined to be $H=H^e+H^A$ with $\{\vert
e_i>,\vert v_j>\}$, i=1, 2, ..., N, and j=1, 2, ..., M, for it's
"in-space" while $\{\vert D_k>, \vert F_l>\}$ for it's "out-space"
with  $1\leq k \leq K $, $1\leq l\leq L $, and N+M=K+L. $H^e$ is the
Hilbert space where the states are defined:
\begin{equation}
\vert\Psi_i>=\sum_{j=1}^N c_{ij}\vert e_j>,
\end{equation}
$H^A$ is the subspace for ancillas, U(N+M) will couple this two
subspace together. Let's use $\omega$  to denote the adjustable
parameter(s) in the unitary transformation, we can define
$\vert\psi>^{out}=U(\omega)\vert \Psi>$ and express it in the
"out-space"
\begin{equation}
\vert\psi_j>^{out}=\sum_{k=1}^{K}d_{jk}(\omega)\vert
D_k>+\sum_{l=1}^{L} f_{jl}(\omega)\vert F_l>,
\end{equation}
with the normalization constraint $\sum_{k=1}^{K}\vert
d_{jk}\vert^2 +\sum_{l=1}^{L}\vert f_{jl}\vert^2=1.$ If we want to
unambiguously discriminate all $\Psi_j$ in G, we should find the
general $U(\omega)$ which gives
\begin{equation}
\vert\psi_j>^{out}=d_{jj}(\omega)\vert D_j>+\sum_{l=1}^{L}
f_{jl}(\omega)\vert F_l>,
\end{equation}
then, $\vert D_j><D_j\vert$ and $\sum_{l=1}^{L}\vert F_l><F_l\vert$
should be the projective operator corresponds to $E_j$ and $E_0$,
respectively. With the $U(\omega)$, we could write , for example,
$\vert D_j>$ in the "in-space" as $\vert D_j>^{in}=U^{-1}\vert
D_j>=\vert D_j>^e+\vert D_j>^A$ with  $\vert D_j>^e$ and $\vert
D_j>^A$ are two non-normalized vectors which lies in $H^e$ and
$H^A$, respectively. There should be
\begin{equation}
E_j(\omega)=\vert D_j>^e<D_j\vert,~~~
E_0(\omega)=\sum_{l=1}^{L}\vert F_l>^e<F_l\vert,
\end{equation}
and $E_0(\omega)+\sum_{j=1}^{N}E_j(\omega)=\textbf{I}^e$. With these
operators in hands, we could get both the optimal operators and
 the maximum values for discriminating $\vert \Psi_j>$.  When the
projective operators are expressed in the "in-space", there are
written formally in terms of  $\vert e_i>< e_j\vert$. If we could
define $\{\vert e_j>\}$ by $\{\vert \Psi_j^{\bot}>\}$ at the
beginning, then we shall be able to complete the task of
expressing $E_m$  in terms of $\vert
\Psi_i^{\bot}><\Psi_j^{\bot}\vert$. The
 argument above can also be generalized to other cases with
 different operations on G.

 In present works, we always adopt the proposals originated from the
 works by Sun ${\sl et~al.}$ [8]: (a) any pure state can be realized by a
 single-photon state and (b), according to Reck's theorem [34], any
 unitary transformation matrix can also be realized by an optical
 network consisting of beam-splitters, phase-shifters, ${\sl
 etc.}$, all these optical elements should construct an one-photon
 interferometer (OPI). The device in Fig.2 is a typical four-port
 beam splitter which is used to realize a two-dimensional unitary
 transformation $U_2(\omega)$:
 \begin{equation}
 \left(%
\begin{array}{cc}
  -\sin\omega & \cos\omega\\
  \cos\omega & \sin\omega \\
\end{array}%
\right)\left(%
\begin{array}{c}
  \vert\omega,r> \\
   \vert\omega,d> \\
\end{array}%
\right)=\left(%
\begin{array}{c}
  \vert\omega,l> \\
   \vert\omega,u> \\
\end{array}%
\right).
\end{equation}
A property of this beam-splitter, which is frequently applied in
present works, should be noted: suppose there is an input
\begin{equation}
\vert\phi>=c_{w,l}\vert\omega,l>+c_{w,u}\vert\omega,u>,
\end{equation}
where $c_{w,l}$ and  $c_{w,u}$ are real parameters for simplicity,
after performing the $U_2(\omega)$, the output should be
\begin{equation}
\vert\phi'>=U_2(\omega)\vert\phi>=c_{w,r}\vert\omega,r>+c
_{w,d}\vert\omega,d>
\end{equation}
with the coefficients satisfy:
\begin{eqnarray}
c_{\omega,r}&=&- c_{\omega,l}\sin\omega+ c_{\omega,u}\cos\omega,\\
c_{\omega,d}&=&\ c_{\omega,l}\cos\omega+ c_{\omega,u}\sin\omega,
\end{eqnarray}
if we choose
\begin{equation}
\sin\omega=\frac{c_{\omega,u}}{ c_{\omega,l}^2+c_{\omega,u}^2},~~
\cos\omega=\frac{c_{\omega,r}}{ c_{\omega,l}^2+c_{\omega,u}^2},
\end{equation}
then $\vert\phi'> = (c_{\omega,l}^2+c_{\omega,u}^2)\vert\omega,d>$
while it's output along $\vert\omega,r>$ is zero.

\begin{figure}
\centering
\includegraphics{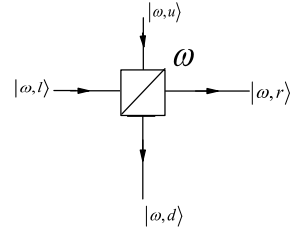}
\caption{\label{fig:epsart} A four-port beam-splitter used to
realize a two-dimensional unitary transformation }
\end{figure}

\subsection{which parameters are known?}
In present work, we shall deal with the case that all the states
in G are linearly independent and their overlaps are also known.

{\sl Definition~~2.1:~} a N-dimensional matrix  O(N) is  defined
by it's matrix elements
\begin{equation}
o_{ij}=<\Psi_i\vert\Psi_j>,
\end{equation}
with constraint that $\vert O_{ij}\vert < 1$ holds for $i\neq j$.

Certainly, O(N) is Hermitian. Using O*(N) and $\tilde{O}(N)$ for
it's  conjugate matrix and transposed matrix, respectively, there
should be $O^*_{ij}=\tilde{O}_{ij}=o_{ji}$.

{\sl Definition~~2.2:~}A(N) is used to denote the adjoint matrix
of O(N), $A_{ij}=a_{ij}$, the inverse of O should be
\begin{equation}
O^{-1}(N)=\frac{A(N)}{\det(O)},
\end{equation}
where $\det(O)$ denotes the determinate of O(N).

{\sl Definition~~2.3:~} $t_j$ is used to denote
\begin{equation}
t_j=<\psi^{\bot}_j\vert\Psi_j>.
\end{equation}

From the definition of the reciprocal states, if $\vert \Psi_j>$
is a the reciprocal state of $\vert \Psi_j>$, then
$\exp\{i\phi_j\}\vert\Psi_j>$ is also a reciprocal state of $\vert
\Psi_j>$. We can always let
$t_j=<\psi^{\bot}_j\vert\Psi_j>=<\psi_j\vert\Psi_j^{\bot}>$ by
choosing a suitable set of  $\{\phi_j\}$. Defining
\begin{equation}
\vert\Psi^{\|}_j>=\frac{1}{\sqrt{1-t_j^2}}(\vert\Psi_j>-t_j\vert\Psi^{\bot}>),
\end{equation}
one may verified that $\vert\Psi_j^{\bot}>$ and
$\vert\Psi^{\|}_j>$ form an orthonormal basis for $\vert\Psi_j>$
and
\begin{equation}
\vert\Psi_j>=\sqrt{1-t_j^2}\vert\Psi^{\|}_j>+t_j\vert\Psi_j^{\bot}>,
\end{equation}
it should be emphasized here that, either $\vert\Psi_j^{\bot}>$ or
$\vert\Psi^{\|}_j>$, is defined from all the states in G:
 \begin{equation}
 {\sl Theorem ~~2.1}:~~~<\Psi_j^{\bot}\vert\Psi_k> = t_j\delta_{jk}.
\end{equation}

With the $t_j$ and O(N) defined above, we may introduced another
transformation matrix:

{\sl Theorem ~~2.2:~~} denoting $(\delta_{ij})=\textbf{I}$, and
defining
\begin{equation}
R(N)=\tilde{O}(N)(\frac{1}{t_j}\delta_{ij}),
\end{equation}
there should be
\begin{equation}
R(N)\vert \Psi^{\bot}>=\vert \Psi>.
\end{equation}
{\sl Proof:~}formally, we can write $\vert\Psi_j>$ as a linear
combination of $\{\vert\Psi^{\bot}_k>\}$ in the way like
$\vert\Psi_j>=\sum_{k=1}^N h_{jk}\vert\Psi^{\bot}_k>$, there should
be $o_{ij}=<\Psi_i\vert\Psi_j>=\sum_{k=1}^N
h_{jk}<\Psi_i\vert\Psi^{\bot}_k>=h_{ji}t_i$, which gives
$\vert\Psi_j>=\sum_{k=1}^N \frac{o_{kj}}{t_k}\vert\Psi^{\bot}_k>$.
Let N=3, as an example, we have
\begin{eqnarray}
\frac{\vert\Psi_1^{\bot}>}{<\Psi_1\vert\Psi_1^{\bot}>}+
\frac{o_{21}\vert\Psi_2^{\bot}>}{<\Psi_2\vert\Psi_2^{\bot}>} +
\frac{o_{31}\vert\Psi_3^{\bot}>}{<\Psi_3\vert\Psi_3^{\bot}>}&=&\vert\Psi_1>\nonumber\\
\frac{o_{12}\vert\Psi_1^{\bot}>}{<\Psi_1\vert\Psi_1^{\bot}>}+
\frac{\vert\Psi_2^{\bot}>}{<\Psi_2\vert\Psi_2^{\bot}>} +
\frac{o_{32}\vert\Psi_3^{\bot}>}{<\Psi_3\vert\Psi_3^{\bot}>}&=&\vert\Psi_2>\nonumber\\
\frac{o_{13}\vert\Psi_1^{\bot}>}{<\Psi_1\vert\Psi_1^{\bot}>}+
\frac{o_{23}\vert\Psi_2^{\bot}>}{<\Psi_2\vert\Psi_2^{\bot}>} +
\frac{\vert\Psi_3^{\bot}>}{<\Psi_3\vert\Psi_3^{\bot}>}&=&\vert\Psi_3>\nonumber\\
\end{eqnarray}
It is possible to express $\vert\Psi^{\bot}>$ in terms of
$\vert\Psi>$ through introducing the inverse of R(N)
\begin{equation}
R^{-1}(N)=(t_{j}\delta_{ij})\frac{\tilde{A}(N)}{\det(O)},
\end{equation}
from Eq.(2.16) and Eq.(2.21) while the relation,
$\det(O)=\det(\tilde{O})$, has been used [33]. Naturally,
\begin{equation}
R^{-1}(N)\vert\Psi>=\vert\Psi^{\bot}>.
\end{equation}
Both R(N) and $R^{-1}$, which are known from O(N), can be used to
derive the value of $t_j$. Let's use N=3, as an example,  to give
the derivation. From Eq.(2.25), we have
\begin{eqnarray}
\frac{t_1}{\det(O(3))}(a_{11}\vert\Psi_1>+a_{21}\vert\Psi_2>
+a_{31}\vert\Psi_3>)&=&\vert\Psi_1^{\bot}>,\nonumber\\
\frac{t_2}{\det(O(3))}(a_{12}\vert\Psi_1>+a_{22}\vert\Psi_2>
+a_{32}\vert\Psi_3>)&=&\vert\Psi_2^{\bot}>,\nonumber\\
\frac{t_3}{\det(O(3))}(a_{13}\vert\Psi_1>+a_{23}\vert\Psi_2>
+a_{33}\vert\Psi_3>)&=&\vert\Psi_3^{\bot}>,\nonumber\\
\end{eqnarray}
times $<\psi_k^{\bot}\vert$ on both sides of each equation, there
are $\frac{t_j^2a_{jj}}{\det(O(3))}=1$ for j=1, 2, 3. This
calculation can be generalized to
\begin{equation}
{\sl Theorem~~ 2.3:~~} t_j=\sqrt{\frac{\det(O(N))}{a_{jj}}}.
\end{equation}

Now,we have shown how to get  $\vert\Psi_j^{\bot}>$ from G, and
their overlaps can be expressed thorough

{\sl Theorem~~2.4:~} defining the matrix $O^{\bot}(N)$ by
\begin{equation}
O^{\bot}_{ij}(N)=o^{\bot}_{ij}=<\Psi_i^{\bot}\vert\Psi^{\bot}_j>,
\end{equation}
there should be
\begin{equation}
o^{\bot}_{ij}=\frac{a_{ij}}{\sqrt{a_{ii}a_{jj}}}.
\end{equation}
{\sl Proof:} we could suppose $\vert\Psi^{\bot}>$ is known at
first while $\vert\Psi>$ can be viewed as it's "reciprocal" state,
and there should be
\begin{equation}
\tilde{O}^{\bot}(\frac{1}{t_j}\delta_{ij})\vert\Psi>=\vert\Psi^{\bot}>
\end{equation}
by following the argument for the case where  $\vert\Psi>$ is known
at first. Comparing it with Eq.(2.25), we find
\begin{equation}
O^{\bot}=(t_{j'}\delta_{i'j'})\frac{{A}}
{\det(O)}(\frac{1}{t_j}\delta_{ij}),
\end{equation}
it can be written in the form of Eq.(2.29) by using Eq.(2.27).
Some $O^{\bot}$ shall be given in the appendix.

\subsection{the double-triangle representation}
A complete set of reciprocal states exists if, and only if, the
state $\vert \psi_j>$ are linearly independent while the reciprocal
states are also linearly independent, this fact will be used in
deriving a set of normalized basis  set $\{\vert e_j>\}$. Letting
\begin{eqnarray}
\vert \Psi^{\bot}_1>&=&\vert e_1>,\\
\vert\Psi^{\bot}_2>&=& o^{\bot}_{12}\vert e_1>+\sqrt{1-\vert
o^{\bot}_{12}\vert^2}\vert e_2>,\nonumber
\end{eqnarray}
the coefficients of $\vert\Psi^{\bot}_2>$ are decided by the two
requirements (a) it's overlap with $\vert\Psi^{\bot}_1>$ keeps
unchanged  and (b) the state should be normalized. These
requirements may also used in deriving the coefficients of
$\vert\Psi_3>$: suppose
\begin{equation}
\vert\Psi^{\bot}_3>=\sum_{k=1}^3 c_{3k}^{\bot}\vert e_k>,
\end{equation}
from the requirements (a) and (b), there are three equations
\begin{eqnarray}
<\Psi^{\bot}_1\vert\Psi^{\bot}_3>&=&c_{31}^{\bot},~~<\Psi^{\bot}_3\vert\Psi^{\bot}_3>=1,\\
<\Psi^{\bot}_2\vert\Psi^{\bot}_3>&=&o_{21}^{\bot}o^{\bot}_{13}+\sqrt{1-\vert
o^{\bot}_{12}\vert^2}c^{\bot}_{32},
\end{eqnarray}
their solutions should be
\begin{eqnarray}
c_{31}^{\bot}&=&o_{13}^{\bot},\nonumber\\
c_{32}^{\bot}&=&\frac{o^{\bot}_{23}-o_{21}^{\bot}o^{\bot}_{13}}{\sqrt{1-\vert o^{\bot}_{12}\vert^2}},\\
c_{33}^{\bot}&=&\sqrt{1-\vert c^{\bot}_{31}\vert^2-\vert
c_{32}^{\bot}\vert^2}.\nonumber
\end{eqnarray}
In principle, this process can be continued until we get all the
coefficients, $c_{ij}^{\bot}$, used as the matrix elements for the
matrix $C^{\bot}(N)$. Introducing another N-dimensional Matrix  E,
which is defined by $E=(C^{\bot})^{-1}$, with it's matrix elements
denoted by $e_{ij}$, we can define the basis, $\{\vert e_j>\}$, in
the way like
\begin{eqnarray}
\vert e_1>&=&\vert\Psi^{\bot}_1>,\nonumber\\
 \vert e_2>&=&\frac{\vert\Psi^{\bot}_2>-o_{12}^{\bot}\vert\Psi^{\bot}_1>}
 {\sqrt{1-\vert o_{12}^{\bot}\vert^2}},\\
\vert e_j>&=&\sum _{i=1}^{j}e_{ji}\vert\Psi_{i}^{\bot}>.\nonumber
\end{eqnarray}
After introducing this basis, every input state can be expressed
in it by defining
\begin{equation}
c_{ij}=<e_j\vert\Psi_i>.
\end{equation}
One may verify that there should  be $c_{ij}=0$ if $i>j$ according
to theorem 2.1, this makes
\begin{equation}
\vert e_N>=\vert \Psi_{N}>
\end{equation}
according to  a simple reasoning, and the expression, $\vert
e_N>=\sum _{i=1}^{N}e_{ji}\vert\Psi_{i}^{\bot}>$, is an equivalent
form of it. The matrix C with $c_{ij}$ is a upper-triangle matrix
while $C^{\bot}$ is a lower-triangle matrix, for  examples,
\begin{equation}
\left(
  \begin{array}{ccc}
    c^{\bot}_{11} & 0 & 0 \\
    c^{\bot}_{21} & c^{\bot}_{22}& 0 \\
    c^{\bot}_{31}& c^{\bot}_{32} & c^{\bot}_{33} \\
  \end{array}
\right)\left(
         \begin{array}{c}
           \vert e_1> \\
           \vert e_2>\\
           \vert e_3>\\
         \end{array}
       \right)=\left(
                 \begin{array}{c}
                   \vert\Psi^{\bot}_1> \\
                   \vert\Psi^{\bot}_2> \\
                   \vert\Psi^{\bot}_3> \\
                 \end{array}
               \right),
\end{equation}
\begin{equation}
\left(
  \begin{array}{ccc}
    c_{11} & c_{12} & c_{13} \\
    0 & c_{22} & c_{23} \\
    0 & 0 & c_{33} \\
  \end{array}
\right)\left(
         \begin{array}{c}
           \vert e_1> \\
           \vert e_2>\\
           \vert e_3>\\
         \end{array}
       \right)=\left(
                 \begin{array}{c}
                   \vert\Psi_1> \\
                   \vert\Psi_2> \\
                   \vert\Psi_3> \\
                 \end{array}
               \right),
\end{equation}
this is the reason why we call  $\{\vert e_j>\}$ the double-triangle
representation (DTR). In the argument below, we always suppose that
the states, either $\{\vert\Psi_i>\}$ or $\{\vert\Psi_j^{\bot}>\}$,
have been expressed in the DTR. At the end of this section, we would
like to emphasis again: if O(N) is known, then $t_j$, $O^{\bot}(N)$,
$\{\vert\Psi^{\bot}_j>\}$, $\{\vert e_i>\}$, C and $C^{\bot}$ are
also given at the same time.

\begin{figure}
\centering
\includegraphics{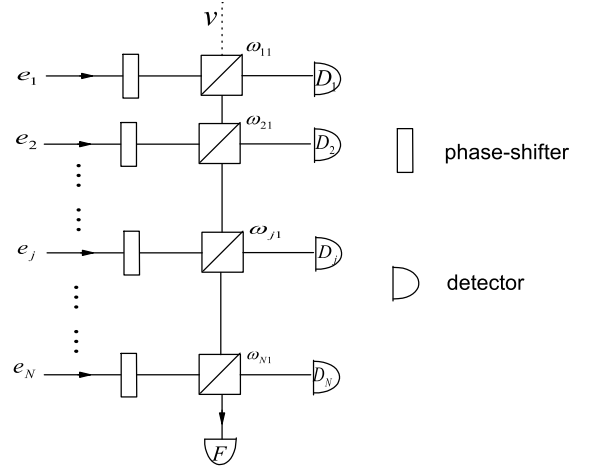}
\caption{\label{fig:epsart} the OPI for filtering  }
\end{figure}
\section{The quantum state filtering}
\subsection{ the POVMs for the filtering}
The quantum state filtering, which was termed  in [21-23], is a
special case of telling whether a state $\vert \Psi_?>$ belongs to ,
$G_1\equiv\{\Psi_1\}$ or $G_2\equiv\{\vert\Psi_k>\}$ for $k\ge 2$,
with a non-zero probability of failure. The derivation of the
optimal measurement strategy, in terms of $\{E_i,E_0\}$ for i=1, 2,
to distinguish $\vert\Psi_1>$ from $G_2$ has been given  and it is
shown that this problem is equivalent to the discrimination of a
pure state and an arbitrary mixed states. The quantum state
filtering, as we shall shown, plays important roles in present
works: (1) it's an excellent example to show how our scheme works
while (2) the filtering  in a successive way will be used to
complete other operations on G. The POVMs $\{E_m\}$ for filtering
are defined by $E_1\vert\Psi_k>=0$ for $k\ge 2$, $E_2\vert\Psi_1>=0$
and $E_0$ for failure, our task is to find an general unitary
transformation $U(\omega)$ which transfers each state $\vert\Psi_j>$
in the "out-space" in way like:
\begin{eqnarray}
\vert \Psi_1>^{out}&=&d_{11}\vert D_1>+ f_1\vert F>,\\
\vert \Psi_k>^{out}&=&\sum_{j=2}^N d_{kj}\vert D_j>+ f_k\vert F>.
\end{eqnarray}
This $U(\omega)$ can be realized by the OPI in Fig.3. Keeping in
mind that $\vert\Psi_k>$ has no input along the rail $e_1$, the
input state should be $\vert\Psi_1>$ when the detector $D_1$ has
been triggered. By applying Eqs.(2.9-14), we are always possible to
prevent the signals of $\vert\Psi_1>$ from appearing in the detector
$ D_k$. Usually, a complex parameter, say, $c_{ij}$ may be expressed
as
\begin{equation}
c_{ij}=\vert c_{ij}\vert\exp\{i\phi_{ij}\},
\end{equation}
with $\exp\{i\phi_{ij}\}=c_{ij}/\vert c_{ij}\vert$. In Fig.3, a
phase-shifter, $\Phi_{ij}$, is placed in front of a beam-splitter
denoted by $\omega_{ij}$, we always choose the phase-shifter
\begin{equation}
\Phi_{j1}=-\phi_{1j},
\end{equation}
while each beam-splitter takes the value
\begin{eqnarray}
\sin\omega_{k1}&=&\frac{\sqrt{\sum_{j=1}^{k-1}\vert c_{1j}\vert
^2-c^2_{11}\sin ^2\omega}}{\sqrt{\sum_{j=1}^{k}\vert
c_{1j}\vert ^2-c^2_{11}\sin ^2\omega}},\\
 \cos\omega_{k1}&=&\frac{\vert c_{1k}\vert}{\sqrt{
\sum_{j=1}^{k}\vert c_{1j}\vert ^2-c^2_{11}\sin ^2\omega}},
\end{eqnarray}
with $\omega_{11}\equiv\omega$ and $c_{ij}=<e_i\vert\Psi>$, for
examples,
\begin{eqnarray}
c_{11}&=&t_1,~~ c_{12}=-\frac{o^{\bot}_{21}t_1}{\sqrt{1-\vert
o^{\bot}_{12}\vert^2}},\\
c_{21}&=&0,~~c_{22}=\frac{t_2}{\sqrt{1-\vert
o^{\bot}_{12}\vert^2}}.
\end{eqnarray}
In Fig.3, we could read
\begin{equation}
\vert D_1>^{in}=\exp\{-i\Phi_{11}\}U^{-1}_2(\omega_{11})\vert
D_1>,
\end{equation}
and get
\begin{equation}
\vert D_1>^{in}=-\sin\omega\vert e_1>+\sin\omega\vert v>,
\end{equation}
where $\Phi_{11}=0$ and Eq.(2.9) have been used. Through a similar
argument, we could arrive at
\begin{eqnarray}
\vert D_2>^{in}&=&\vert D_2>^e+\cos\omega_{21}\sin\omega\vert
v>,\\
\vert D_2>^{e}&=&-\sin\omega_{21} e^{-i\Phi_{21}}\vert
e_2>+\cos\omega_{21}\cos\omega\vert e_1>,\nonumber
\end{eqnarray}
it can be proved that
\begin{equation}
\vert D_2>^e=\frac{-\cos\omega
e^{i\phi_{12}}}{\sqrt{\cos^2\omega+\sin^2\omega\vert
o^{\bot}_{12}\vert^2}}\vert\Psi_2^{\bot}>.
\end{equation}
In fact, we may use the relation, $E_2(\omega)=\textbf{I
}^e-E_1({\omega})-E_0(\omega)$, instead of giving all $\vert
D_k>^{e}$ in detail. Using Eq.(3.1) and Eq.(3.11), we may get
\begin{eqnarray}
\vert D_1>^e&=&-\sin\omega\vert\Psi^{\bot}_1>,\\
 \vert F>^e&=&\frac{(\textbf{I}^e-\vert
D_1>^e<D_1\vert )\vert\Psi_1>}{\sqrt{<\Psi_1\vert
(\textbf{I}^e-\vert D_1>^e<D_1\vert )\vert\Psi_1>}},\nonumber
\end{eqnarray}
and the POVMs of filtering should be \begin{eqnarray}
E_1(\omega)&=&\vert D_1>^e<D_1\vert,~E_0(\omega)=\vert
F>^e<F\vert,\\
 E_2(\omega)&=&\sum_{k=2}^N\vert
 D_k>^e<D_k\vert=\textbf{I}^e-E_1(\omega)-E_0(\omega).\nonumber
 \end{eqnarray}
If the POVMs $\{E_m\}$ were known, then the calculation of the
optimal value of filtering should be  easily completed. Suppose
$\eta_j$ is the ${\sl a~ priori}$ probability of $\vert\Psi_j>$,
we denote $P_{1/(N-1)}(\omega)$ and $F_{1/(N_1)}(\omega)$ the
average value of success and failure of filtering, respectively,
\begin{eqnarray}
P_{1/(N-1)}(\omega)&=&\sum_{i=1}^2 Tr(E_i\rho_i),\nonumber\\
F_{1/(N-1)}(\omega)&=& Tr(E_0\rho),\\
 P_{1/(N-1)}(\omega)&+&F_{1/(N-1)}(\omega)=\sum_{j=1}^N\eta_j,
 \end{eqnarray}
with $\rho_1=\eta_1\vert\Psi_1><\psi_1\vert$,
$\rho_2=\sum_{k=2}^N\eta_k\vert\Psi_k><\Psi_k\vert$ and
$\rho=\rho_1+\rho_2$. A simple calculation shows that
\begin{equation}
F_{1/(N-1)}(\omega)=\eta_1\vert f_1\vert^2+\sum_{k=2}^N\eta_k\vert
f_k\vert^2,
\end{equation}
with $f_j=<F\vert\psi_j>$ and
\begin{equation}
f_1=\sqrt{1-t_1^2\sin^2\omega},~f_k=\frac{o_{1k}}{\sqrt{1-t_1^2\sin^2\omega}}.
\end{equation}
The optimal value of $F_{1/(N-1)}(\omega)$,
$F^{opt}_{1/(N-1)}(\omega)=F_{1/(N-1)}(\omega^{opt})$, is defined
to be minimum value of $F_{1/(N-1)}(\omega)$ in the domain of
$0\le \omega<2\pi$. From Eqs.(3.17-18), there is
\begin{equation}
\frac{dF_{1/(N-1)}(\omega)}{d\omega}=t_1^2\sin
2\omega(\eta_1-\frac{\sum_{k=2}\eta_k\vert
o_{1k}\vert^2}{1-t_1^2\sin^2\omega}),
\end{equation}
and $F^{opt}_{1/(N-1)}(\omega)$ happens at
$dF_{1/(N-1)}(\omega)/d\omega\vert_{\omega=\omega^{opt}}=0.$ Now,
we are able to give the optimal values of filtering: (a) if
$\sqrt{\sum_{k=2}^N \frac{\eta_k}{\eta_1}\vert o_{1k}\vert^2}>1$,
by letting $\sin^2\omega^{opt}=0$, we have
\begin{equation}
F^{opt}_{1/(N-1)}(\omega)=\eta_1+\sum_{k=1}^{N}\vert
O_{1k}\vert^2,
\end{equation}
(b) if $1-t_1^2\le\sqrt{\sum_{k=2}^N \frac{\eta_k}{\eta_1}\vert
o_{1k}\vert^2}\le 1$, through letting
\begin{equation}
\sin\omega^{opt}=\frac{1-\sqrt{\sum_{k=2}^N
\frac{\eta_k}{\eta_1}\vert o_{1k}\vert^2}}{t_1^2},
\end{equation}
we arrive at
\begin{equation}
F^{opt}_{1/(N-1)}(\omega)=2\sqrt{\eta_1\sum_{k=2}^N \eta_k\vert
o_{1k}\vert^2},
\end{equation}
and (c) if $\sqrt{\sum_{k=2}^N \frac{\eta_k}{\eta_1}\vert
o_{1k}\vert^2}<1-t_1^2$, the optimal value should be
\begin{equation}
F^{opt}_{1/(N-1)}(\omega)=\eta_1(1-t_1^2)+\frac{\sum_{k=2}^N\eta_k\vert
o_{1k}\vert^2}{1-t_1^2},
\end{equation}
while $\sin^2\omega^{opt}=1.$ Substituting $\omega^{opt}$ for
$\omega$ in Eqs.(3.13-14), we could also get the optimal POVMs
 $\{E^{opt}_m\}$.
\subsection{an example: filtering for N=3}
 The filtering  of N=3 is a case with a fully analytical solution
 and an optical implementation of the optimal strategy [23], we
 shall show, via a simple optical setting,  how to recover all the optimal values in
 [23]. For N=3, there is
\begin{equation}
1-t_1^2=\frac{\vert o_{12}\vert^2+\vert
 o_{13}\vert^2-o_{12}o_{23}o_{31}-o_{21}o_{13}o_{32}}{1-\vert o_{23}\vert^2}.
\end{equation}
Defining
\begin{equation}
A=\eta_2\vert o_{12}\vert^2+\eta_3\vert o_{13}\vert^2,
\end{equation}
we can write the general results of filtering to the  N=3: (a) for
$A>\eta_1$, there is
\begin{equation}
F^{opt}_{1/2}=\eta_1+A,
\end{equation}
(b) for $\eta_1(1-t_1^2)^2\le A\le \eta_1$, there should be
\begin{equation}
F^{opt}_{1/2}=2\sqrt{\eta_1 A},
\end{equation}
and (c) else, $A<\eta_1(1-t_1^2)^2$,
\begin{equation}
F^{opt}_{1/2}=\eta_1(1-t_1^2)+\frac{A}{1-t_1^2}.
\end{equation}
Besides all this optimal results, we could also get the optimal
POVMs for filtering with N=3.

\begin{figure}
\centering
\includegraphics{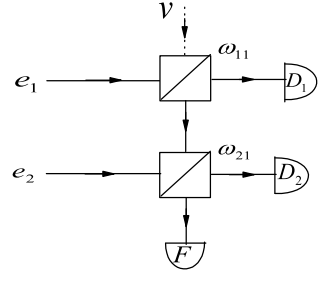}
\caption{\label{fig:epsart} the OPI for filtering with N=2 }
\end{figure}

\subsection{filtering with N=2:~discriminating two pure states}
The problem, how to discriminate $\vert\Psi_1>$ from $\vert\Psi_2>$,
is the most interesting case in the field of quantum states
unambiguously discrimination. Here, it can be solved as a special
case of filtering with N=2. The present solution is complete in the
sense that: not only the optimal values but also the optimal POVMs
should be given at the same time while the POVMs have the forms
required by Eqs.(2.2-4). The OPI in Fig.4 is for the filtering with
N=2 and it takes $<\Psi_1\vert\Psi_2>=<\Psi_2\vert\Psi_1>$ for
simplicity. From the DTR for N=2, the basis vectors should be
\begin{equation}
\vert e_1>=\vert\Psi_1^{\bot}>,~~\vert e_2>=\vert \Psi_2>,
\end{equation}
and the states have the forms
\begin{eqnarray}
\vert\Psi_1>&=&\sqrt{1-\vert o_{12}\vert^2}\vert e_1>+o_{21}\vert e_2>,\nonumber\\
\vert\Psi_2>&=&\vert e_2>.
\end{eqnarray}
With known parameters for N=2, which have been given in the
Appendix, we have
\begin{eqnarray}
\vert
\psi_1^{\bot}>&=&\frac{\vert\Psi_1>-o_{21}\vert\Psi_2>}{\sqrt{1-\vert
o_{12}\vert^2}},\nonumber \\
\vert\psi_2^{\bot}>&=&\frac{\vert\Psi_2>-o_{12}\vert\Psi_1>}{\sqrt{1-\vert
o_{12}\vert^2}},
\end{eqnarray}
by applying Eq.(2.25).  The POVMs for discriminating two linearly
independent states, $\vert\Psi_1>$ and $\vert\Psi_1>$, should be
\begin{eqnarray}
E_1&=&\sin^2\omega\vert\Psi^{\bot}_1><\psi_1^{\bot}\vert,\nonumber \\
E_2&=&\frac{\cos^2\omega}{\cos^2\omega+\sin^2\omega\vert
o_{12}\vert^2}\vert\Psi^{\bot}_2><\psi_2^{\bot}\vert,\\
E_0&=&\frac{(\vert\Psi_1>-t_1^2\sin^2\omega\vert\Psi^{\bot}_1>)(
<\Psi_1\vert-t_1^2\sin^2\omega<\Psi^{\bot}_1\vert)}{1-t^2_1\sin^2\omega}\nonumber
\end{eqnarray}
while $t_1=\sqrt{1-\vert o_{12}\vert^2}$, and
\begin{equation}
F_{1/1}(\omega)=\eta_1(1-t_1^2\sin^2\omega)+\frac{\eta_2\vert
o_{12}\vert^2}{1-t_1^2\sin^2\omega}.
\end{equation}
Now, we could give the optimal values and the optimal POVMs at the
same time: (1) for $\sqrt{\frac{\eta_2}{\eta_1}}>\frac{1}{\vert
o_{12}\vert}$, let $\sin^2\omega^{opt}=0$, there should be
\begin{eqnarray}
E_1^{opt}&=&0,~~E_2^{opt}=\vert\Psi_2^{\bot}><\Psi^{\bot}_2\vert,\nonumber\\
E_0^{opt}&=&\vert\Psi_1><\Psi_1\vert,\nonumber\\
F_{1/1}^{opt}(\omega)&=&\eta_1+\eta_2\vert o_{12}\vert^2,\\
P_{1/1}^{opt}(\omega)&=&\eta_2(1-\vert o_{12}\vert^2),\nonumber
\end{eqnarray}
(2) if $\vert
o_{12}\vert\le\sqrt{\frac{\eta_2}{\eta_1}}\le\frac{1}{\vert
o_{12}\vert}$, by letting
\begin{equation}
\sin^2\omega^{opt}=\frac{1-\sqrt{\frac{\eta_2}{\eta_1}}\vert
o_{12}\vert} {1-\vert o_{12}\vert^2},
\end{equation}
we shall get the optimal POVMs
\begin{eqnarray}
E_1^{opt}&=&\frac{1-\sqrt{\frac{\eta_2}{\eta_1}}\vert o_{12}\vert}
{1-\vert o_{12}\vert^2}\vert
\Psi^{\bot}_1><\Psi_1^{\bot}\vert,\nonumber\\
E_2^{opt}&=&\frac{1-\sqrt{\frac{\eta_1}{\eta_2}}\vert o_{12}\vert}
{1-\vert o_{12}\vert^2}\vert
\Psi^{\bot}_2><\Psi_2^{\bot}\vert,\nonumber\\
E_0^{opt}&=&\textbf{I}^e-\sum_{j=1}^2 E_j^{opt},
\end{eqnarray}
which give the optimal results
\begin{eqnarray}
F_{1/1}^{opt}(\omega)&=&2\sqrt{\eta_1\eta_2}\vert o_{12}\vert,\\
P_{1/1}^{opt}(\omega)&=&\sum_{j=1}^2\eta_j-2\sqrt{\eta_1\eta_2}\vert
o_{12}\vert,\nonumber
\end{eqnarray}
and (3) when $\sqrt{\frac{\eta_2}{\eta_1}}<\vert o_{12}\vert,$,
through choosing $\sin^2\omega^{opt}=1$, we arrive at
\begin{equation}
E_1^{opt}=\vert\Psi_1^{\bot}><\Psi_1^{\bot}\vert,~E_0^{opt}=\vert\Psi_2><\Psi_2\vert,
\end{equation}
while $E_2^{opt}=0$, all these operators leads to
\begin{eqnarray}
F_{1/1}^{opt}(\omega)&=&\eta_1\vert o_{12}\vert^2+\eta_2,\\
P_{1/1}^{opt}(\omega)&=&\eta_1(1-\vert o_{12}\vert^2).\nonumber
\end{eqnarray}
Compare with other methods of solving the same question, the
present scheme states that the general POVMs may be given before
the decision of the optimal values of success and failure.
\section{filtering in subspace}
\subsection{filtering with the background}
\begin{figure}
\centering
\includegraphics{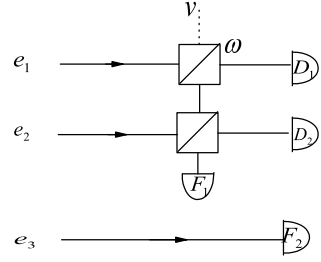}
\caption{\label{fig:epsart} The OPI for  discriminating two states
with the background }
\end{figure}
Suppose $G_1\equiv\{\vert\Psi_k>\}$ and
$G_2\equiv\{\vert\Psi_l>\},$ a new operation of G can be specified
by the definition of the POVMs as $E_1\vert\Psi_?>=0$ if
$\vert\Psi_?>$ belongs to $G_2$,
 $E_2\vert\Psi_?>=0$ if $\vert\Psi_?>$ belongs to $G_1$ and $E_0$
corresponds to failure. If a state, say, $\vert\Psi_g>$, is shared
by both $G_1$ and $G_2$, then $E_1\vert\Psi_g>=E_2\vert\Psi_g>=0$
 according to the definition of the POVMs $\{E_m\}$. We call this
case the name of discriminating with the background. In this
section, we shall consider a simple case of discriminating
$\vert\Psi_1>$ from $\vert\Psi_2>$ with $\vert\Psi_3>$ as the
background. In the DTR, this operation on G can also be viewed as a
filtering in a two-dimensional subspace.

For N=3, the the basis vectors in the DTR should be:
\begin{eqnarray}
\vert e_1>&=&\vert\Psi_1^{\bot}>,\nonumber\\
\vert
e_2>&=&\frac{\vert\Psi_2^{\bot}>-o^{\bot}_{12}\vert\Psi_1^{\bot}>}{\sqrt
{1-\vert o_{12}^{\bot}\vert^2}},\\
\vert e_3>&=&\vert\Psi_3>,\nonumber
\end{eqnarray}
with
\begin{equation}o_{12}^{\bot}=\frac{o_{13}o_{32}-o_{12}}{\sqrt{(1-\vert
o_{23}\vert^2)(1-\vert o_{13}\vert^2)}},
\end{equation}
while the matrix C(3) takes the form
\begin{equation}
\left(%
\begin{array}{ccc}
  t_1 & \frac{o_{12}-o_{13}o_{32}}{\sqrt{1-\vert o_{32}\vert^2}}& o_{31} \\
  0 & \sqrt{1-\vert o_{32}\vert^2} & o_{32} \\
  0 & 0& 1 \\
\end{array}%
\right),
\end{equation}
with $t_1$, which holds  for N=3, is
\begin{equation}
t_1=\sqrt{\frac{\det(O(3))}{1-\vert o_{23}\vert^2}}.
\end{equation}
The $U(\omega)$ in Fig.5 is required to transform each
$\vert\Psi_j>$ to $\vert\Psi_j>^{out}$ as
\begin{eqnarray}
\vert\Psi_1>^{out}&=&d_1\vert D_1>+f_{11}\vert F_1>+f_{12}\vert
F_2>,\nonumber\\
\vert\Psi_2>^{out}&=&d_2\vert D_2>+f_{21}\vert F_1>+f_{22}\vert
F_2>,\\ \vert\Psi_3>^{out}&=&\vert F_2>,\nonumber
\end{eqnarray}
and this goal can be reached, as we shall shown later, by applying
the argument of  filtering. The $U^{-1}(\omega)$ shall give
\begin{eqnarray}
\vert D_1>^e&=&-\sin\omega \vert \Psi^{\bot}_1>,\nonumber\\
\vert F_2>^e&=&\vert\Psi_3>,\\
\vert F_1>^e&=&\frac{(I^e-\vert F_2>^e<F_2\vert-\vert
 D_1>^e<D_1\vert)\vert\Psi_1>}{\sqrt{1-\vert f_{12}\vert^2-\vert d_1\vert^2}},\nonumber
\end{eqnarray}
with  $\vert  D_2>^{e}$ has given by Eq.(3.12). The POVMs are
defined by
\begin{eqnarray}
E_1(\omega)&=&\vert D_1>^e<D_1\vert,~~E_2(\omega)=\vert
D_2>^e<D_2\vert,\nonumber\\
E_0(\omega)&=&\vert F_1>^e<F_1\vert+\vert F_1>^e<F_1\vert.
\end{eqnarray}
Defining
\begin{eqnarray}
F_1(\omega)&=&\sum_{j=1}^2\eta_j\vert f_{j1}\vert^2,\\
F_2(\omega)&=&\sum_{i=1}^3\eta_i\vert f_{i2}\vert^2,
\end{eqnarray}
with $f_{ij}=<F_j\vert\Psi_i>$, the average value of failure
should be
\begin{equation}
F(\omega)=F_1(\omega)+F_2(\omega).
\end{equation}
A simple calculation shows
\begin{equation}
F_2(\omega)=\sum_{j=1}^2\eta_j\vert o_{3j}\vert^2+\eta_3,
\end{equation}
it is still in a special form of filtering with N=3, see Eq.(3.26).
The $ F_1(\omega)$ is left to be decided by the general results of
filtering with N=2.

Formally, $\vert\Psi_1>$ and $\vert\Psi_1>$ can written by
\begin{eqnarray}
\vert\Psi_1>&=&\sqrt{1-\vert c_{13}\vert ^2}\vert
\Psi_1'>+c_{13}\vert e_3>,\nonumber\\
\vert\Psi_2>&=&\sqrt{1-\vert c_{23}\vert ^2}\vert
\Psi_2'>+c_{23}\vert e_3>,
\end{eqnarray}
with $\vert\Psi_1'>$ and $\vert\Psi_2'>$ are two normalized states
defined in the subspace specified by $\{\vert e_j>\}$ for j=1,2:
\begin{eqnarray}
\vert\Psi_1'>&=&\frac{c_{11}\vert e_1>+c_{12}\vert
e_2>}{\sqrt{1-\vert c_{13}\vert^2}},\nonumber\\
\vert\Psi_2'>&=&\vert e_2>,
\end{eqnarray}
and their overlap should be
\begin{equation}
o'_{12}=<\Psi_1'\vert\Psi_2'>=\frac{o_{12}-o_{13}o_{32}}{\sqrt{(1-\vert
o_{31}\vert^2)(1-\vert o_{32}\vert^2)}}.
\end{equation}
Now, in the two-dimensional subspace with $\{\vert e_1>,\vert
e_2>\}$, our task is to discriminate $\vert \Psi_1>$ from $\vert
\Psi_2>$ with
\begin{equation}
\eta'_j=\eta_j(1-\vert o_{3j}\vert^2),
\end{equation}
for j=1,2, to be their ${\sl a ~priori}$ probability, respectively.
 According to our discussion about filtering, we have
\begin{eqnarray}
 F_1(\omega)&=&\eta'_1(1-(1-\vert
 o_{12}'\vert^2)\sin^2\omega)\nonumber\\
 &&+\frac{\eta_2'\vert o_{12}'\vert^2}{1-(1-\vert
 o_{12}'\vert^2)\sin^2\omega},
\end{eqnarray}
which is equivalent with the one given by Eq.(4.9). This is the
reason why the present case is viewed as a process of filtering
with N=2, certainly, in the subspace without the background. It's
optimal results have nearly the same forms for filtering with N=2:
 (1) for $\sqrt{\frac{\eta'_2}{\eta'_1}}>\frac{1}{\vert
o'_{12}\vert}$, let $\sin^2\omega^{opt}=0$, there should be
\begin{eqnarray}
F^{opt}(\omega)&=&F_2(\omega)+\eta'_1+\eta'_2\vert o'_{12}\vert^2,\\
P^{opt}(\omega)&=&\eta'_2(1-\vert o'_{12}\vert^2),\nonumber
\end{eqnarray}
(2) if $\vert
o'_{12}\vert\le\sqrt{\frac{\eta'_2}{\eta'_1}}\le\frac{1}{\vert
o'_{12}\vert}$, by letting
\begin{equation}
\sin^2\omega^{opt}=\frac{1-\sqrt{\frac{\eta'_2}{\eta'_1}}\vert
o'_{12}\vert} {1-\vert o'_{12}\vert^2},
\end{equation}
we shall get the optimal  results
\begin{eqnarray}
F^{opt}(\omega)&=&F_2(\omega)+2\sqrt{\eta'_1\eta'_2}\vert o'_{12}\vert,\\
P^{opt}(\omega)&=&\sum_{j=1}^2\eta'_j-2\sqrt{\eta'_1\eta'_2}\vert
o'_{12}\vert,\nonumber
\end{eqnarray}
and (3) when $\sqrt{\frac{\eta'_2}{\eta'_1}}<\vert o'_{12}\vert,$
through choosing $\sin^2\omega^{opt}=1$, we arrive at
\begin{eqnarray}
F^{opt}(\omega)&=&F_2(\omega)+\eta'_1\vert o'_{12}\vert^2+\eta_2,\\
P^{opt}(\omega)&=&\eta'_1(1-\vert o'_{12}\vert^2).\nonumber
\end{eqnarray}
It should be noted that $F_2(\omega)$, which has been given in
Eq.(4.11), is in fact a constant. The  present argument, which is
suitable for  discriminating
$G_1\equiv\{\vert\psi_1>,\vert\Psi_3>\}$ and
$G_2\equiv\{\vert\psi_2>,\vert\Psi_3>\}$, can be generalized to
the
 discriminating two general mixtures sharing part of states in
comm.
\subsection{discriminating two mixtures in Jordan basis}
\begin{figure}
\centering
\includegraphics{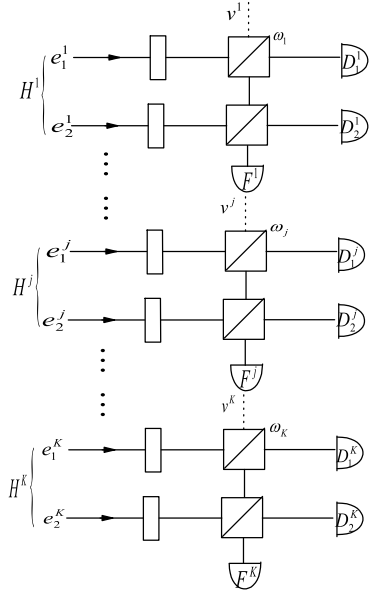}
\caption{\label{fig:epsart} The OPI for filtering the mixtures in
Jordan basis }
\end{figure}
Suppose there are two mixtures,
\begin{eqnarray}
\rho_1&=&\sum_{i=1}^K p_i\vert\Psi_i><\Psi_i\vert,\nonumber\\
\rho_2&=&\sum_{j=K+1}^{2K} p_j\vert\Psi_j><\Psi_j\vert,
\end{eqnarray}
with $\sum_{i=1}^K p_i=\sum_{j=K+1}^{2K} p_j=1$. Let $\lambda_k$ to
be the ${\sl a~priori}$ probability for $\rho_k$, k=1, 2, we may
introduce $\eta_i=\lambda_1 p_i$ as the ${\sl a~priori}$ probability
for $\vert\Psi_i>$ in $G_1\equiv\{\vert\Psi_i>\}$ while
$\eta_j=\lambda_2 p_j$ as the ${\sl a~priori}$ probability for
$\vert\Psi_i>$ in $G_2\equiv\{\vert\Psi_j>\}$. If $<
\Psi_k\vert\Psi_{k'}>=0$ for $k\neq k'$ except
$<\Psi_i\vert\Psi_{i+k}>=\cos\theta_i$, $\rho_1$ and $\rho_2$ are
called in Jordan basis. Defining the POVMs:
$E_1\vert\Psi_j>=E_2\vert\Psi_i>=0$ while $E_0$ for failure, this
$\{E_m\}$ can be get, as it has been show by the works in [21-23],
through discriminating pairs of pure states in each subspace.

The OPI in Fig.5 is used to discriminate these two mixtures,
$\rho_1$ and $\rho_2$. The total Hilbert space here is defined by
$H=\sum_{i=1}^K\oplus H^i$, each $H^i$ is a two-dimensional subspace
with it's basis as
\begin{eqnarray}
\vert e^i_1>&=&\frac{\vert\Psi_i>-\cos\theta_i\vert\Psi_{K+i}>}{\sin\theta_i},\nonumber\\
\vert e^i_2>&=&\vert\Psi_{K+i}>,
\end{eqnarray}
the two states in this $H^e$ are
\begin{eqnarray}
\vert \Psi_i>&=&\sin\theta_i\vert e^i_1>+\cos\theta_i\vert
e^i_2>,\nonumber\\
\vert \Psi_{K+i}>&=&\vert e^i_2>,
\end{eqnarray}
while their reciprocal states
\begin{eqnarray}
\vert
\Psi^{\bot}_i>&=&\frac{\vert\Psi_i>-\cos\theta_i\vert\Psi_{K+i}>}
{\sin\theta_i},\nonumber\\
\vert
\Psi^{\bot}_{K+i}>&=&\frac{\vert\Psi_{K+i}>-\cos\theta_i\vert\Psi_i>}
{\sin\theta_i}.
\end{eqnarray}
In this $H^i$, our task is to filter $\vert \Psi_i>$ from $\vert
\Psi_{K+i}>$, the POVMs, $\{E^i_m\}$, to complete this task should
be
\begin{eqnarray}
E_1^i(\omega_i)&=&\sin^2\omega_i\vert\Psi^{\bot}_i><\Psi^{\bot}_i\vert,\nonumber,\\
E_2^i(\omega_i)&=&\frac{\cos^2\omega_i}{\cos^2\omega_i+\sin^2\omega_i\cos^2\theta_i}
\vert\Psi^{\bot}_{K+i}]><\Psi^{\bot}_{K+i}\vert,\nonumber \\
E_0^i(\omega_i)&=&\textbf{I}^i-\sum_{k=1}^2 E^i_k(\omega_i),
\end{eqnarray}
the average value of the failure in $H^i$ should be
\begin{eqnarray}
F^i(\omega_i)&=&\eta_i(1-\sin^2\theta_i\sin^2\omega_i)\nonumber\\
&&+\frac{\eta_{K+i}\cos^2\theta_i}{1-\sin^2\theta_i\sin^2\omega_i},
\end{eqnarray}
it's  optical values are given by the theorem of  filtering with
N=2. Finally, we can define the POVMs $\{E_m\}$ by
\begin{equation}
E_m=\sum_{i=1}^K E^i_m,
\end{equation}
for m=0, 1, 2. The average value of fail can be expressed by
\begin{equation}
F(\omega)=\sum_{i=1}^K F^i(\omega_i),
\end{equation}
while it's optimal value
\begin{equation}
F^{opt}(\omega)=\sum_{i=1}^K F^i(\omega_i=\omega_i^{opt}),
\end{equation}
where  $\omega_i^{opt}$ should depend on the actual value of the
parameters, $\eta_i$, $\eta_{K+i}$ and $\cos\theta_i$, here, this
requirement has also been pointed
 by the  recent work [20].

\begin{figure}
\centering
\includegraphics{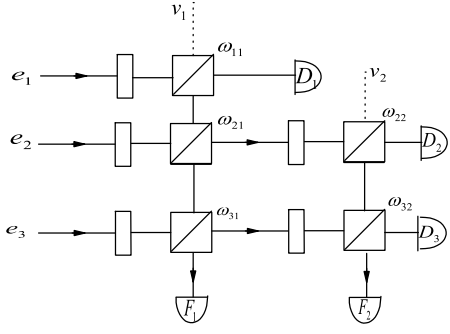}
\caption{\label{fig:epsart} the OPI for discriminating three pure
stats }
\end{figure}
\section{The successive filtering for discrimination of pure states}
\subsection{the optical realization of $U(\omega)$}

In present section, we shall show the POVMs, which are defined for
discriminating of pure states, can be realized in an enlarged
Hilbert space by applying the successive filtering. The OPI in
Fig.7 is designed to discriminate three linearly independent
states,
 $\vert \Psi_i>$ for i=1, 2, 3, here. The $U(\omega)$ realized by this OPI
 can be written as
 \begin{equation}
 U(\omega)=R(2)R(1),
 \end{equation}
with R(1) denotes the unitary transformation done by the
beam-splitters and phase-shifters, $\omega_{j1}$ and $\Phi_{j1}$ for
j=1, 2, 3, on the left part of Fig.7, while R(2) denotes the unitary
transformation realizes by, $\omega_{k1}$ and $\Phi_{k1}$ for k=2,
3, the beam-splitters and phase-shifters on the right part. At
first, the R(1) is defined to filter $\vert\Psi_1>$ from the states,
 $\vert\Psi_2>$ and $\vert\Psi_3>$, here.
\begin{eqnarray}
R(1)\vert\Psi_1>&=&d_{11}\vert D_1>+f_{11}\vert F_1>,\\
R(1)\vert\Psi_k>&=&\sqrt{1-\vert
f_{k1}\vert^2}\vert\Psi'_k>+f_{k1}\vert F_1>,\nonumber
\end{eqnarray}
for k=2, 3, while  $\vert \Psi_k'>$ lies in a two-dimensional
subspace H', which is specified by it's basis as
 $\{\vert\omega_{k1},r>\}$ for k=2, 3, and
\begin{eqnarray}
\vert \Psi_2'>&=& \sqrt{1-\vert o'_{23}\vert^2}
\vert\omega_{21},r>+o'_{32}\vert \Psi_3'>,\nonumber\\
\vert \Psi_3'>&=&e^{i\phi_3'}\vert\omega_{31},r>,
\end{eqnarray}
where $o_{23}'=<\Psi_2'\vert\Psi'_3>$, using Eq.(5.2), we get
\begin{equation}
o'_{23}=\frac{o_{23}-f^*_{21}f_{31}}{\sqrt{(1-\vert f_{21}\vert^2)
(1-\vert f_{31}\vert^2)}},
\end{equation}
 In this run of filtering, the average value of fail should be
\begin{equation}
F_1(\omega)=\sum_{j=1}^3\eta_j\vert f_{j1}\vert^2,
\end{equation}
with $f_{j1}=<F_1\vert\Psi_j>$. After the first turn of filtering,
we are left with two states, $\vert\Psi'_2>$ and $\vert\Psi'_3>$ in
H', with their ${\sl a ~priori}$ probabilities to be
\begin{equation}
\eta'_k=\eta_k(1-\vert f_{k1}\vert^2),
\end{equation}
for k=2, 3, respectively. R(2) is designed to filter $\vert\Psi'_2>$
from $\vert\Psi'_3>$ in the way like:
\begin{eqnarray}
R(2)\vert\Psi_2'>&=&\frac{1}{\sqrt{1-\vert
f_{21}\vert^2}}(d_{22}\vert D_2>+f_{22}\vert F_2>),\nonumber\\
R(2)\vert\Psi_3'>&=&\frac{1}{\sqrt{1-\vert
f_{31}\vert^2}}(d_{33}\vert D_3>+f_{32}\vert F_2>),\nonumber\\
\end{eqnarray}
and in this run of filtering, the average value of the failure
should be
\begin{eqnarray}
F_2(\omega)&=&\eta_2'(1-(1-\vert
o_{23}'\vert^2)\sin^2\omega_2)\nonumber\\
&&+\frac{\eta_3'\vert o_{23}'\vert^2}{1-(1-\vert
o_{23}'\vert^2)\sin^2\omega_2},
\end{eqnarray}
with $\omega_{22}\equiv\omega_2$. If
\begin{equation}
\eta_2=\eta_3=\eta,~~\vert f_{21}\vert=\vert f_{31}\vert,
\end{equation} then we shall get
\begin{equation}
F_2^{opt}=2\eta\vert o_{23}-f^*_{21}f_{31}\vert,
\end{equation}
by letting
\begin{equation}
\sin^2\omega_2^{opt} =\frac{1-\vert o'_{23}\vert}{1-\vert
o'_{23}\vert^2}.
\end{equation}
\subsection{the POVMs for discriminating three pure states}
 By performing the  R(2) after R(1), the state $\vert\Psi_j>$ are
 transformed into:
 \begin{eqnarray}
 \vert\Psi_1>^{out}&=&d_{11}\vert D_1>+f_{11}\vert
 F_1>,\nonumber\\
\vert\Psi_2>^{out}&=&d_{22}\vert D_2>+f_{21}\vert
 F_1>+f_{22}\vert F_2>,\\
\vert\Psi_3>^{out}&=&d_{33}\vert D_3>+f_{31}\vert
 F_1>+f_{32}\vert F_2>. \nonumber
 \end{eqnarray}
 With $U^{-1}(\omega)=R^{-1}(1)R^{-1}(2)$, we can arrive at
 \begin{eqnarray}
\vert D_1>^e&=&-\sin\omega_1\vert\psi^{\bot}_1>,\nonumber\\
\vert
D_2>^e&=&\frac{\cos\omega_1\sin\omega_2}{\sqrt{\cos\omega_1^2+\sin^2\omega_1
\vert o_{12}^{\bot}\vert ^2}}\vert\Psi_2^{\bot}>,\nonumber\\
\vert
F_1>^e&=&\frac{\vert\Psi_1>-t_1\sin^2\omega\vert\Psi^{\bot}_1>}{\sqrt{1-t_1^2\sin^2\omega_1}},\\
\vert F_2>^e&=&\frac{(\textbf{I}^e-\vert D_2>^e<D_2\vert-\vert
F_1>^e<F_1\vert)\vert\Psi_2>}{\sqrt{1-\vert d_{22}\vert^2-\vert
f_{21}\vert^2}}\nonumber,
\end{eqnarray}
with $d_{ij}=<D_j\vert\Psi_i>$, $f_{ij}=<F_j\vert\Psi_i>$ and
$\omega_1=\omega_{11}$.  For discriminating three pure states, the
POVMs are defined by
\begin{eqnarray}
E_1(\omega)&=&\vert D_1>^e<D_1\vert,\nonumber\\
E_2(\omega)&=&\vert D_2>^e<D_2\vert,\\
E_0(\omega)&=&\sum_{k=1}^2\vert F_k>^e<F_k\vert,\nonumber\\
E_3(\omega)&=&\textbf{I}^e-\sum_{l=1}^2E_l-E_0,\nonumber
\end{eqnarray}
the average value of failure is defined by
\begin{equation}
F(\omega)=\sum_{j=1}^3 \eta_j\vert<\Psi_j\vert F_1>\vert^2+
\sum_{l=1}^2 \eta_l\vert<\Psi_l\vert F_2>\vert^2,
\end{equation}
it can be proved that
\begin{equation}
F(\omega)=F_1(\omega)+F_2(\omega),
\end{equation}
the  $F_2(\omega)$, which has been  given in  Eq.(5.8), is in the
form of filtering with N=2. With calculations that
\begin{eqnarray}
f_{11}&=&\sqrt{1-t_1^2\sin\omega_1},\nonumber\\
f_{k1}&=&=\frac{o_{1k}}{\sqrt{1-t_1^2\sin\omega_1}},
\end{eqnarray}
for k=1,2, $F_1(\omega)$ should be
\begin{equation}
F_1(\omega)=\eta_1(1-t_1^2\sin\omega_1)+\frac{\sum_{k=2}^3\eta_k\vert
o_{1k}\vert^2}{1-t_1^2\sin\omega_1},
\end{equation}
certainly, it is also in a typical form of filtering with N=3.
\subsection{the analytic optimal results for a special case}
    Usually, it is difficult for us to give an analytic solution
for  the optimal values of the $F(\omega)$ in Eq.(5.16), while the
following case, which has been discussed in [8], is an exception.
Considering the case, where
$<\Psi_1\vert\Psi_2>=<\Psi_1\vert\Psi_2>=s_1$ and
$<\Psi_2\vert\Psi_3>=s_2$ under the conditions that $s_1^2<s_2$, we
find that
\begin{eqnarray}
t_1&=&\sqrt{\frac{\det(O(3)}{1-s_2^2}},\nonumber\\
 \det(O(3))&=&1-2s_1^2-s_2^2+2s_1^2s_2,
\end{eqnarray}
according to the results given in the appendix. Suppose
$\eta_j=\frac{1}{3}$ for j=1, 2, 3, we could find
\begin{equation}
f_{21}=f_{31}=\frac{s_1}{\sqrt{1-t_1^2\sin\omega_1}}.
\end{equation}
Using Eq.(5.10), there should be
\begin{equation}
F_2^{opt}(\omega_1)=\frac{2}{3}\vert
s_2-\frac{s_1^2}{1-t_1^2\sin^2\omega_1}\vert,
\end{equation}
the optimal value is  defined as the minimum value of the function
\begin{equation}
F(\omega)=F_1(\omega_1)+ F_2^{opt}(\omega_1),
\end{equation}
and it should depend on the actual situations about the $o_{ij}$:
(1) if $2s_1\le \frac{s_1^2}{s_2}$, by letting
\begin{equation}
\sin^2\omega_1^{opt}=\frac{1-2s_1}{1-\frac{2s_1^2}{1+s_2}},
\end{equation}
we shall  get the optimal result
\begin{equation}
F^{opt}=\frac{2}{3}(2s_1-s_2),
\end{equation}
and (2), is $s_1\le 2s_2$, the optimal value
\begin{equation}
F^{opt}=\frac{1}{3}(\frac{s_1^2}{s_2}+2s_2),
\end{equation}
with the choice of
\begin{equation}
\sin^2\omega_1^{opt}=\frac{1-\frac{s_1^2}{s_2}}{1-\frac{2s_1^2}{1+s_2}},
\end{equation}
substituting $\omega_1^{opt}$ for $\omega_1$ in Eq.(5.11), which is
$\omega_1$-dependent, we can get the actual optimal setting for
$\omega_2^{opt}$. One check that:  the optimal values for
$F(\omega)$, which have been given in Eqs.(5.24-25), are consistent
with the optimal results in [8].
\section{the successive filtering for discrimination of two
mixtures}
\subsection{the optical realization in the enlarged space.}
Suppose there are two mixtures
\begin{equation}
\rho_1=\sum_{i=1}^2 P_i \vert\Psi_i><\Psi_i\vert,~~
\rho_2=\sum_{j=3}^4 P_j\vert\Psi_j><\Psi_j\vert,~~
\end{equation}
with $\sum_{i=1}^2 P_i=\sum_{j=3}^4 P_j=1$, each $\rho_i$ with
it's ${\sl a~priori}$ probability to be $\lambda_i$, and
$\sum_{i=1}^2\lambda_i=1$. Letting $\eta_i=\lambda_1\cdot P_i$ for
i=1,2, and $\eta_j=\lambda_2\cdot P_j$ for j=3, 4, the above
question can also be viewed as an operation on
$G\equiv\{\vert\Psi_k>\}$ with $\eta_k$ as it probability for k=1,
2, 3, 4, we are required to tell if a state $\vert\Psi_?>\in G$
belongs to $G_1\equiv\{\vert\Psi_1>,\vert\Psi_2>\}$ or
$G_2\equiv\{\vert\Psi_3>,\vert\Psi_4>\}$ while there is a
non-vanishing probability for fail. In terms of POVMs,
 $E_1\vert\psi_j>=E_2\vert\Psi_i>=0$ holds for  $\vert\Psi_j>\in G_2$ and
$\vert\Psi_i>\in G_1$, respectively. Certainly, there is
$E_0=\textbf{I}^e-\sum_{k=1}^2E_k$. The OPI in Fig.8 is designed to
realize the unitary transformation in the way like
\begin{equation}
U(\omega)=R(2)R(1),
\end{equation}
where R(1) is the unitary transformation for filtering
$\vert\Psi_2>$ from the rest of the states in G, R(2) is  used to
filter $\vert\Psi'_1>$ from $\vert\Psi'_3>$ and $\vert\Psi'_4>$,
whose definitions shall be given later. It should be noted that,
when the detector $D_1$ fired, we can not tell whether this signal
is from $\vert\Psi_1>$ or $\vert\Psi_2>$ since the fact that these
two states may have non-zero coefficients, $c_{12}$ and $c_{22}$,
along the rail $e_2$, respectively. It is certain that this signal
can not come from the states, $\vert\Psi_3>$ and  $\vert\Psi_4>$,
 according to our discussion of DTR. One may compare the present
$R(1)\vert\Psi_j>$,
 with the one in discriminating
 three pure states,
\begin{eqnarray}
R(1)\vert\Psi_2>&=& d_{21}\vert D_1>+f_{21}\vert F_1>,\nonumber\\
R(1)\vert\Psi_1>&=&\sqrt{1-\vert f_{11}\vert^2-\vert
d_{11}\vert^2}\vert\Psi'_1>\nonumber\\&&+ d_{11}\vert
D_1>+f_{11}\vert F_1>,\\
R(1)\vert\Psi_k>&=&\sqrt{1-\vert f_{k1}\vert^2}\vert\Psi'_k>+
f_{k1}\vert F_1>,\nonumber
\end{eqnarray}
for k=3,4. Defining a Hilbert space H' with it's basis as
$\{\vert\omega_{21},r>,\vert\omega_{31},r>,\vert e_1>\}$, there are
three states:
\begin{eqnarray}
\vert\Psi'_1>&=&\frac{c_{11}}{\sqrt{1-\vert f_{11}\vert^2-\vert
d_{11}\vert^2}}\vert e_1>\nonumber\\
&&+\frac{o'_{31}-o'_{34}o'_{41}}{\sqrt{1-\vert o'_{43}\vert^2}}\vert
\omega_{21},r>+o'_{41}\vert \Psi'_4>\nonumber\\
\vert\Psi'_3>&=&\sqrt{1-\vert o'_{43}\vert^2}\vert \omega_{21},r>
+o'_{43}\vert \Psi'_4> ,\\
\vert\Psi'_4>&=&e^{i\phi'_4}\vert \omega_{31},r>,\nonumber
\end{eqnarray}
with $o'_{ij}=<\Psi'_i\vert\Psi'_j>$,  these overlaps can be derived
from Eq.(6.3) by the requirement that the $o_{ij}$ keeps unchanged
when the unitary transformation is performed on the input,
\begin{equation}
o'_{1k}=\frac{o_{1k}-f^*_{11}f_{k1}}{\sqrt{(1-\vert f_{11}\vert^2
-\vert d_{11}\vert^2)(1-\vert f_{k1}\vert^2)}},
\end{equation}
for k=3,4, while
\begin{equation}
o'_{34}=\frac{o_{34}-f_{31}^*f_{41}}{\sqrt{(1-\vert f_{31}\vert^2)
(1-\vert f_{41}\vert^2)}}.
\end{equation}
In principle, we could realize R(2) as the filtering for N=3,
\begin{eqnarray}
R(2)\vert\Psi'_1>&=&\frac {d_{12}\vert D_1>+ f_{12}\vert
F_2>}{\sqrt{(1-\vert f_{11}\vert^2 -\vert
d_{11}\vert^2)}},\nonumber\\
R(2)\vert\Psi'_k>&=&\frac{\sum_{l=3}^4 d_{kl}\vert
D_l>+f_{k2}\vert F_2>}{\sqrt{1-\vert f_{k1}\vert^2 }}.
\end{eqnarray}
In the first run of filtering, the average value of failure is
defined by
\begin{equation}
F_1(\omega)=\sum_{j=1}^{N=4}\eta_j\vert f_{j1}\vert^2,
\end{equation}
while the one for the second run of filtering is
\begin{eqnarray}
F_2(\omega)&=&\eta_1'(1-c_{11}'\sin^2\omega_2)  \nonumber\\
&&+\frac{\sum_{k=3}^4\eta_k'\vert o'_{1k}\vert^2}
{1-c_{11}'\sin^2\omega_2},
\end{eqnarray}
 in a standard form of filtering  with N=3, and their ${\sl a~ priori}$ probability
is
\begin{eqnarray}
\eta'_1=\eta_1(1-\vert f_{11}\vert^2 -\vert
d_{11}\vert^2),\nonumber\\
\eta'_k=\eta_k(1-\vert f_{k1}\vert^2 ),
\end{eqnarray}
for k=3, 4, respectively. The coefficient, $ c_{11}'$, may be read
from Eq.(6.4) as
\begin{equation}
c_{11}'=\frac {c_{11}}{\sqrt{(1-\vert f_{11}\vert^2 -\vert
d_{11}\vert^2)}}.
\end{equation}
\begin{figure}
\centering
\includegraphics{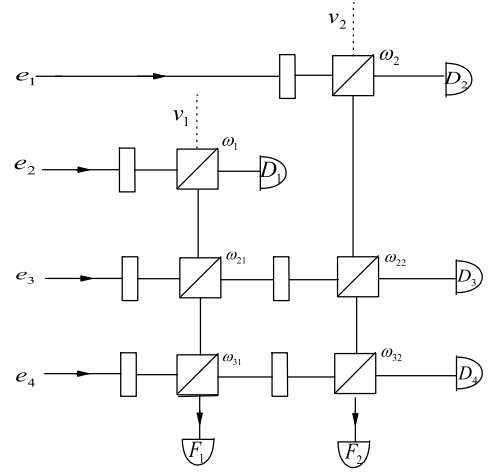}
\caption{\label{fig:epsart} The OPI for discriminating two
mixtures }
\end{figure}

\subsection{the POVMs realized by the OPI in Fig.8}
As we have shown, it's possible for us to get a $U(\omega)$ which
transforms each $\vert\Psi_j>$ in the out-space:
\begin{eqnarray}
\vert \Psi_2>^{out}&=&d_{21}\vert D_1>+f_{21}\vert F_1>,\nonumber\\
\vert\Psi_1>^{out} &=&\sum_{j=1}^2 d_{1j}\vert D_j>+\sum_{i=1}^2
f_{1i}\vert F_{1i}>,\\
\vert\Psi_k>^{out}&=&\sum_{l=3}^4 d_{kl}\vert D_l>+\sum_{i=1}^2
f_{ki}\vert F_i>.\nonumber
\end{eqnarray}
Using these expressions and the inverse $R^{-1}(1)R^{-1}(2)$, we can
get
\begin{eqnarray}
\vert D_1>^e&=&-\sin\omega_1\vert e_2>,\nonumber\\
\vert F_1>^e&=&\frac{(\textbf{I}^e-\vert D_1>^e<
D_1\vert)\vert\Psi_2>}{\sqrt{1-\vert d_{21}\vert^2}},\nonumber\\
 \vert
D_2>^e&=&-\sin\omega_2\vert e_1>,\\
\vert F_2>^e&=&\frac{(\textbf{I}^e-\sum_{i=1}^2\vert
D_i>^e<D_i\vert-\vert F_1>^e< F_1\vert)\vert \Psi_1>} {\sqrt{1-\vert
f_{11}\vert^2 -\sum_{i=1}^2\vert d_{1i}\vert^2)}} \nonumber
\end{eqnarray}
with $ d_{ij}=<D_j\vert\Psi_i>$ and $ f_{ij}=<F_j\vert\Psi_i>$, the
POVMs are defined by
\begin{eqnarray}
E_1&=&\sum_{i=1}^2\vert D_i>^e<D_i\vert,\nonumber\\
E_0&=&\sum_{i=1}^2\vert F_i>^e<F_i\vert,\\
E_2&=&\sum_{j=3}^4\vert
D_j>^e<D_j\vert=\textbf{I}^e-\sum_{m=0}^1E_m,\nonumber
\end{eqnarray}
With these operators in hands, we could define the average value of
fail
\begin{equation}
F(\omega)=\sum_{j=1}^{N=4}\eta_j<\Psi_j\vert E_0\vert\Psi_j>,
\end{equation}
and one may check that
\begin{equation}
F(\omega)=F_1(\omega)+F_2(\omega).
 \end{equation}
 The coefficients, which  are needed in the calculation of $F(\omega)$,
are list here
\begin{eqnarray}
f_{11}&=&\frac{o_{21}-c_{12}c_{22}\sin^2\omega_1}{\sqrt{1-c_{22}^2\sin^2\omega_1}},\nonumber\\
f_{21}&=&\sqrt{1-c_{22}^2\sin^2\omega_1},\nonumber\\
f_{k1}&=&\frac{o_{2k}}{\sqrt{1-
c_{22}^2\sin^2\omega_1}},\nonumber\\
d_{11}&=&-c_{12}\sin\omega_1,\\
d_{12}&=&-c_{11}\sin\omega_2,\nonumber\\
f_{12}&=&\sqrt{1-\sum_{i=1}^2\vert d_{1i}\vert^2-\vert
f_{11}\vert^2},\nonumber\\
f_{k2}&=&\frac{o_{1k}-f^*_{11}f_{k1}}{\sqrt{1-\sum_{i=1}^2\vert
d_{1i}\vert^2-\vert f_{11}\vert^2}},\nonumber
\end{eqnarray}
with k=3, 4.
\begin{figure}
\begin{center}
\includegraphics[scale=0.5]{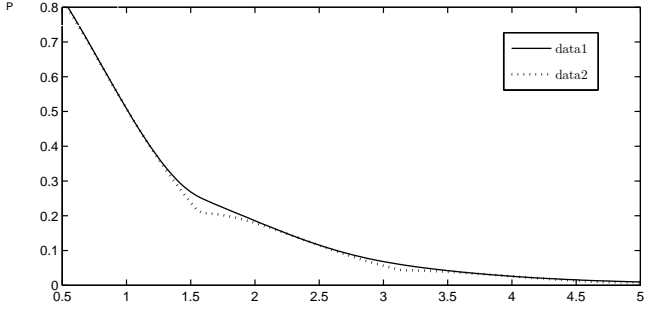}
\end{center}
\caption{\label{fig:epsart}the optimal results of $F^{opt}$ in
Eq.(6.16), the present results are described by the solid line,
while the analytical results in dash line.}
\end{figure}

\begin{figure}
\begin{center}
\includegraphics[scale=0.5]{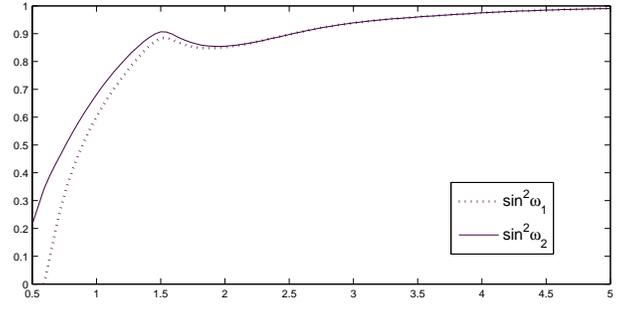}
\end{center}
\caption{\label{fig:epsart}the optimal choice of $\omega$}
\end{figure}

\subsection{an application of the POVMs}
In a recent work, Raynal ${\sl et~ al.}$ considered the question,
which came from the implementation of the BB84 by using the four
quantum optical coherent states  $\{\vert \pm \alpha>$,$\vert\pm
 i \alpha>\}$ [35], of how to discriminate the following two mixtures:
\begin{eqnarray}
\rho_1&=&\frac{1}{2}(\vert\alpha><\alpha\vert+\vert-\alpha><-\alpha\vert),\nonumber\\
\rho_2&=&\frac{1}{2}(\vert
i\alpha><i\alpha\vert+\vert-i\alpha><-i\alpha\vert),
\end{eqnarray}
and the authors expressed the optimal failure probability in terms
of the mean photon number:
 \begin{equation}
 F^{opt}=e^{-\mu}(\vert\cos \mu\vert+\vert\sin\mu\vert),
\end{equation}
with $\mu\equiv\vert\alpha\vert^2$.

Here, we shall reconsider this problem with the POVMs in
Eqs.(13-14). Writing the all the known parameters in terms of
$\alpha$, we designed a program to get the optimal velue of
$F(\omega)$ in Eq.(6.15) by scanning in the parameters space $0\le
\omega_1,\omega_2\le 2\pi.$ The final result of our calculation and
the analytic solution in Eq.(6.19) are both presented in Fig.9,
while the optimal values of the $\omega_1$ and $\omega_2$ are given
in Fig.10. Although in some regions  of Fig.9, small discrepancy
still exists, our numerical calculations are consistent with the
analytical solutions well in most parts of the  parameter space,
this fact shall give great supports to our present proposal.

\begin{figure*}
\begin{center}
\includegraphics{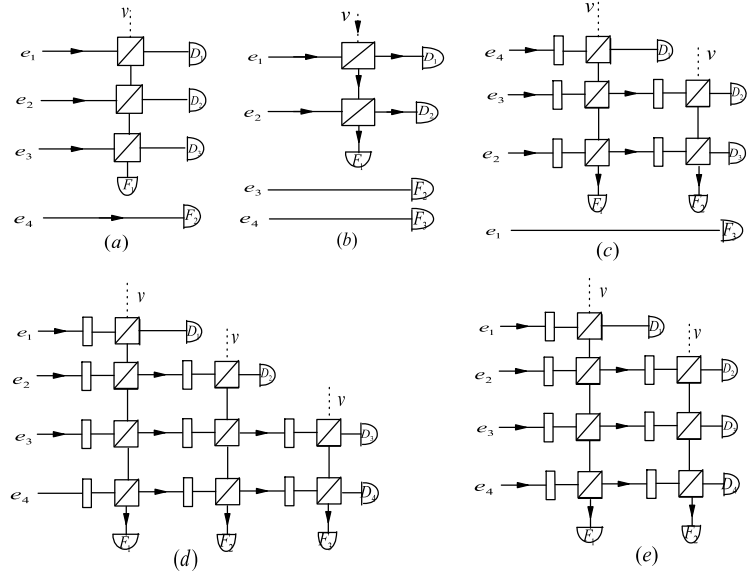}
\end{center}
\caption{\label{fig:epsart}The OPIs for the following cases: (a)
filtering $\vert\Psi_1>$ from $\vert\Psi_2>$ and $\vert\Psi_3>$
with $\vert\Psi_4>$ as the background; (b) filtering
$\vert\Psi_1>$ from $\vert\Psi_2>$ with $\vert\Psi_3>$ and
$\vert\Psi_4>$ as the background; (c) discriminating three pure
states with $\vert\Psi_4>$ as the background; (d) discriminating
four pure states and (e) discriminating $\vert\Psi_1>$,
$\vert\Psi_2>$ and $\{\vert\Psi_3>,\vert\Psi_4>\}$. }
\end{figure*}

\section{discussion}
    In the present paper, we always adopt a naive
 understanding of the mixture: suppose  a mixtures is denoted
by $\rho$, for examples, $\rho=\sum_{j=1}^N
\eta_j\vert\Psi_j><\Psi_j\vert$,  in each run of the experiment,
 the input for our OPI is still a pure state belonging to the
set $\{ \vert\Psi_j>\}$.

 For the system of N=2 and N=3, we have given a series of
derivations to show why these cases can be solved by applying the
argument of filtering, certainly, within the DTR. The  case
discussed in Sec.VI, is an important case of N=4. There are still
others types of operations for N=4 system: (a) filtering
$\vert\Psi_1>$ from $\vert\Psi_2>$ and $\vert\Psi_3>$ with
$\vert\Psi_4>$ as the background; (b) filtering $\vert\Psi_1>$
from $\vert\Psi_2>$ with  $\vert\Psi_3>$ and  $\vert\Psi_4>$ as
the background; (c) discriminating three pure states with
$\vert\Psi_4>$ as the background; (d) discriminating four pure
states and (e) discriminating $\vert\Psi_1>$, $\vert\Psi_2>$ and
$\{\vert\Psi_3>,\vert\Psi_4>\}$. Their OPIs are given in Fig.11.
In principle, all these cases can be solved by reducing to
filtering. For the cases where  more than once filtering is
needed, it's hard for us to find analytical solutions for the
optical values.

A important profit of our scheme should  be mentioned here:  the
POVMs for each case shall be able to, although maybe not in a
 optimal way,  complete the task of discriminating
when the \emph{a priori} probability of each state is not
completely decided.

In end of this paper, we would like to emphasize that : first, for a
given case, if one could prepare the input in the one-photon state,
then our OPI can be directly used for the optical experimental
realization. Then, although the POVMs are from the one-photon
picture, yet  they are are general and state-type independent.
Finally, a proposal, rather than a complete proof, has been given
here in order to find a solution to the problem of the quantum state
unambiguous discrimination. It's still a open question that: whether
the task of quantum state unambiguous discriminating, either of pure
states or of mixtures,  can be solved by reducing it  to the problem
of quantum state filtering?

\begin{acknowledgments}
We wish to acknowledge the helpful discussion of Prof. Chen L.X.
\end{acknowledgments}

\appendix*
\section{some
  known matrices for the
low-dimensional cases.}
 For N=2,
\begin{equation}
O(2)=\left(%
\begin{array}{cc}
  1 & o_{12} \\
  o_{21} & 1 \\
\end{array}%
\right),
\end{equation}
\begin{equation}
O^{\bot}(2)=\left(%
\begin{array}{cc}
  1 & -o_{12} \\
  -o_{21} &1 \\
\end{array}%
\right),
\end{equation}
with
\begin{equation}
\det(O^{-1}(2))=1-\vert o_{12}\vert ^2,~~t_1=t_2=\sqrt{1-\vert
o_{12}\vert ^2}
\end{equation}
For N=3,
\begin{equation}
O(3)=\left(%
\begin{array}{ccc}
  1 & o_{12} & o_{13}\\
  o_{21}& 1 & o_{23}\\
  o_{31}& o_{32} & 1\\
\end{array}%
\right),
\end{equation}
with A(3), the adjoint of O(3), to be
\begin{equation}
\left(%
\begin{array}{ccc}
  1-\vert o_{23}\vert^2 & o_{13}o_{32}-o_{12} & o_{12}o_{23}-o_{13} \\
  o_{23}o_{31}-o_{21} & 1-\vert o_{13}\vert^2& o_{21}o_{13}-o_{23}\\
  o_{32}o_{21}-o_{31}& o_{31}o_{12}-o_{32} & 1-\vert o_{12}\vert^2\\
\end{array},%
\right)
\end{equation}
The inverse $O_{-1}(3)$ could be given by
$O_{-1}(3)=\tilde{A}(3)/\det(O)$ with
\begin{eqnarray}
\det(O(3))=&&1-\vert o_{12}\vert^2-\vert o_{13}\vert^2-\vert
o_{23}\vert^2\nonumber\\&&+o_{12}o_{23}o_{31}+o_{21}o_{13}o_{32}
\end{eqnarray}
which has applications in
\begin{equation}
t_1=\sqrt{\frac{\det(O)}{1-\vert o_{23}\vert^2}},\\
t_2=\sqrt{\frac{\det(O)}{1-\vert o_{13}\vert^2}},\\
t_3=\sqrt{\frac{\det(O)}{1-\vert o_{21}\vert^2}}. \end{equation}
The overlap matrix , $O^{\bot}(3)$, is known with the form:
\begin{widetext}\begin{equation}
O^{\bot}(3)=\left(%
\begin{array}{ccc}
  1& \frac{o_{13}o_{32}-o_{12}}{\sqrt{(1-\vert o_{23}\vert^2)
  (1-\vert o_{13}\vert^2)}}& \frac{o_{12}o_{23}-o_{13}}{\sqrt{(1-\vert o_{23}\vert^2)
  (1-\vert o_{12})}}\\
  \frac{o_{23}o_{31}-o_{21}}{\sqrt{(1-\vert o_{23}\vert^2)
  (1-\vert o_{13}\vert^2)}} & 1& \frac{o_{21}o_{13}-o_{23}}{\sqrt{(1-\vert o_{13}\vert^2)
  (1-\vert o_{12}\vert^2)}}\\
  \frac{o_{32}o_{21}-o_{31}}{\sqrt{(1-\vert o_{23}\vert^2)
  (1-\vert o_{12}\vert^2)}}& \frac{o_{31}o_{12}-o_{32}}{\sqrt{(1-\vert o_{13}\vert^2)
  (1-\vert o_{12}\vert^2)}} & 1\\
\end{array},%
\right)
\end{equation}
\end{widetext}

\end{document}